\documentstyle[preprint,eqsecnum,prl,epsf,aps]{revtex}

\begin{document}
\draft
\preprint{}
\title
{PARTON DISTRIBUTIONS IN NUCLEON ON THE BASIS OF 
 A RELATIVISTIC INDEPENDENT  QUARK MODEL}
\author
{N.Barik}
\address
{Department of Physics, Utkal University,VaniVihar,
 Bhubaneswar-751004,India} 
 \author
 {R.N.Mishra}
 \address
 {Department of Physics,
 Dhenkanal college,
 Dhenkanal-759001,Orissa,India.}
 
\maketitle

\begin{abstract}

At a low  resolution scale with $Q^2={\mu}^2$ corresponding to the
nucleon bound state; deep inelastic unpolarized structure functions 
$F_1(x,{\mu}^2)$ and $F_2(x,{\mu}^2)$ are derived with correct 
support using the symmetric part of the hadronic tensor under
some simplifying assumptions in the Bjorken limit. For 
doing this; the nucleon in its ground state has been represented
by a suitably constructed momentum wave packet of its valence 
quarks in their appropriate SU(6) spin flavor configuration 
with the momentum probability amplitude taken phenomenologically
in reference to the independent quark model of scalar-vector
harmonic potential. The valence quark distribution functions
$u_v(x,{\mu}^2)$ and $d_v(x,{\mu}^2)$, extracted from the 
structure function $F_1(x,{\mu}^2)$ in a parton model 
interpretation, satisfy normalization constraints as well 
as the momentum sum-rule requirements at a bound state scale 
of ${\mu}^2=0.1 GeV^2$. QCD evolution of these distribution 
functions taken as the inputs; yields at $Q_0^2=15 GeV^2;
xu_v(x,Q_0^2)$ and $xd_v(x,Q_0^2)$ in good qualitative 
agreement with the experimental data. The gluon distribution
$G(x,Q_0^2)$ and the sea-quark distribution $q_s(x,Q_0^2)$;
which are dynamically generated using the leading order
renormalization group equation; also match reasonably 
well with the available experimental data. 

\end{abstract}


\section{Introduction}

It is well known that low-energy description of hadron structure
in terms of constituent quark models 
have been quite successful in explaining 
a large body of relevant experimental data. But at very high energies, 
quantum chromodynamics (QCD), the theory of strong interactions of quarks
and gluons; sets a different framework of a more complex quark-parton
picture of hadrons for understanding the deep-inelastic 
scattering(DIS) phenomena. In this picture the deep-inelastic
lepton-nucleon scattering is described
in terms of unpolarized structure functions $F_1(x,Q^2)$ and $F_2(x,Q^2)$;
which are expressed as the charge squared weighted combinations of 
quark-parton distribution functions $f_{\wp}(x,Q^2)$. These parton
distribution functions $f_{\wp}(x,Q^2)$; interpreted as the probability of 
finding a parton ${\wp}$(quark or gluon) in the hadron with a fraction
`x' of the hadron momentum when probed with very high momentum transfer
$Q^2$; play an important role in the standard model phenomenology
providing a deeper understanding of the quark gluon structure of the
hadron at very high energies. In this connection many experiments 
have been made to measure the deep-inelastic structure functions from
which parton distributions inside the nucleon at very high energy have
been extracted [1,2]. Although $Q^2$-dependence of the parton distribution 
functions(PDF) is successfully described by Dokshitzer-Gribov-Lipatov-
Altarelli-Parisi(DGLAP) evolution equations [3] within perterbative
QCD; absolute values of these observables are not provided theoretically 
by QCD to be compared with the experimental data. This is because; it requires
some initial input distribution at lower resolution scale $Q^2={\mu}^2$
which has not been possible from a first principle QCD-calculation due to 
the inadequate understanding of the non-perturbative QCD in the 
confinement domain. Although lattice QCD as a favourite first 
principle technique has been pursued in this context [4]; it 
does involve inevitably increasing computational complexity in arriving
at any desirable precision in its prediction. Therefore it had been
a common practice to take the initial input distributions at a lower 
reference scale in suitable parametrized forms; which are fitted 
ultimately after the QCD evolution with the available experimental data.
Alternatively, there has been attempts to derive the distribution 
functions at the bound state scales of the nucleons described by the
low energy QCD inspired phenomenological constituent quark models;
 which has been pursued over the years by many authors [5-13] with
the purpose of establishing a much desired link between the low
energy constituent quark picture and the high energy quark-parton picture 
of the hadron structure which may provide better understanding of the
parton distribution in nucleons inside the nucleus as well as of the 
parton contributions to the proton spin.
\par The structure functions derivable from a constituent quark model
corresponding to a low energy resolution scale $Q^2={\mu}^2{\simeq}
{\cal O}({\Lambda}_{QCD}^2)$; is considered to represent the twist 
two non-singlet part of the physical structure function. Since at 
higher $Q^2$-region; it is the twist two part of the physical 
structure function that dominates; QCD evolution of the model-derivable
structure functions at $Q^2={\mu}^2$ can provide results for comparison
with the available data at higher-$Q^2$. However the structure functions 
and the parton distributions derived at the bound state scale in 
constituent quark models usually encounter a pathological problem
by not vanishing beyond $x=1$ as required by energy-momentum 
conservation; which is commonly described as  `poor support'. Basing
on the study of one-dimensional Bag model; Jaffe [14] had suggested
a mapping of the distribution function so derived from the region
$0\; {\leq}\; x \;{\leq}\; \infty$, to the kinematically allowed region
$0\; {\leq}\; x \;{\leq}\; 1$, which was applied to three dimensions as well,  
for removing the support problem. However this was just a prescription
only. The problem has been addressed in the center of mass Bag model [7];
where an effective co-variant electromagnetic current of the nucleon 
is considered which satisfies the translational invariance and hence 
conserves the four momentum. Another approach of using the Pierels-
Yoccoz projection was also suggested by Benesh and Miller [6].
Calculation based on Bethe-Salpter and light cone formalism [11] 
do avoid the support problem. Bickerstaff and Londergan [12] have tried
with a different picture of the nucleon; where the confined constituent
quarks are treated approximately as a system of infinite free fermion 
gas at finite temperature. Most of these early calculations with or
without the support problem; yield more or less qualitatively reasonable
results by way of fitting the experimental data with the QCD evolved 
structure functions or the parton distributions realized from the model
input expressions.
\par In our earlier work,we also attempted to derive the 
structure functions of the nucleon at a 
low resolution scale in an alternative constituent  quark model
of relativistic independent quarks confined by an effective 
scalar-vector harmonic potential in a Dirac formalism; whose 
model parameters had been fixed earlier at the level of 
hadron spectroscopy and static hadron properties [15]. The predictive
power of this model had also been successfully demonstrated in wide
ranging low energy hadronic phenomena which include the weak and 
electromagnetic decays of light and heavy flavor mesons [16], elastic
form factors and charge radii of nucleon  [15],  pion and kaon [17] and 
the electromagnetic polarizability of proton [18].  
Extending this model to the study of deep-inelastic scattering 
of electrons off a nucleon; we had obtained quite encouraging 
predictions for the polarized structure functions $g_1^p(x,Q^2)$
and $g_2^p(x,Q^2)$ [19] as well as the unpolarized structure functions
$F_2^p(x,Q^2)$,$F_2^n(x,Q^2)$ with the resulting parton distributions
[20] at a qualitative level. In these works we had taken the usual 
approximation that the nucleon at some static point of $Q^2={\mu}^2$;
consists only of the valence quarks with no gluons or sea-quarks
as constituents. The model solutions for the bound valence quark
eigen-modes provide the essential model input in expressing the
electromagnetic currents which ultimately define the relevant
hadronic tensor for deep-inelastic process. Explicit functional 
forms of the polarized as well as unpolarized structure functions 
were then derived analytically from the antisymmetric and symmetric
part respectively of the hadronic tensor in the Bjorken limit.
However the structure functions so derived at the model scale expectedly
encountered the support problem, although it was found to be minimal.
Therefore in the present work; we would like to improve upon our
earlier attempts by a somewhat different approach within the scope of
the same model in order to realize correct support in the structure
function from which the parton distributions in the nucleon can be extracted.

\par For doing this; we describe the nucleon in its ground state
by a suitably constructed momentum wave packet of its  
valence quarks in appropriate SU(6) spin-flavor configuration;
where each of these quarks is taken in its respective momentum states
with a momentum probability
amplitude derivable from its bound state energy eigen-mode 
obtained in the model. The wave packet includes explicitly a
four delta function to ensure energy momentum conservation at
the composite level. The quark field operators defining the 
electromagnetic currents in the hadronic tensor are expressed as 
free field expansions. Then the unpolarized structure functions 
$F_1^p(x,{\mu}^2)$; derived from the symmetric part of the hadronic
tensor with certain simplifying assumptions in the Bjorken limit,
is found to be free from the support problem. It becomes also
true for the valence quark distributions extracted from the structure
function after appropriate comparison with its parton model 
interpretation, which furthermore satisfy the
normalization requirements as well as 
momentum sum-rule constraints at the bound state scale. We therefore 
believe that these valence quark distributions can provide adequate
model based inputs for QCD-evolution to experimentally relevant
higher $Q^2$-region for a meaningful comparison with the 
experimental data.
\par The paper is organized in the following manner. In sec-II; we
discuss briefly the basic formalism with necessary model inputs
to describe the nucleon in its ground state as a wavepacket
conserving energy-momentum from its constituent level of the
three valence quarks taken in their respective definite momentum
states with appropriate momentum probability amplitudes 
corresponding to their ground state eigen-modes. In sec-III we 
derive the unpolarized structure functions $F_1(x,Q^2)$ and
$F_2(x,Q^2)$ for the nucleon from the symmetric part of the 
hadronic tensor under certain siimplifying assumptions in the 
Bjorken limit. Sec-IV provides an appropriate parton-model
interpretation of $F_1(x,Q^2)$ leading to the extraction 
of the valence quarks distribution functions $u_v(x,Q^2)$ and
$d_v(x,Q^2)$ at a model scale of low $Q^2={\mu}^2$. These
valence distribution functions are found to satisfy the required 
normalization constraints. The bound state scale of $Q^2={\mu}^2$,
which is not explicitly manifested in the expressions for the 
distribution functions, is fixed on the basis of the renormalization
group equations [13] by taking the experimental data of the momentum
carried by the valence quarks at $Q_0^2=15GeV^2$ along with the
same at $Q^2={\mu}^2$. The valence distribution functions 
$u_v(x,Q^2),\;d_v(x,Q^2)$ are evolved to the higher reference
scale $Q_0^2=15GeV^2$ using the QCD non-singlet evolution 
equations; from which valence contributions to the structure 
functions such as $[F_2^p(x,Q_0^2)]_v$,$F_2^n(x,Q_0^2)]_v$
and the combination $[F_2^p(x,Q_0^2)-F_2^n(x,Q_0^2)]_v=
{\frac{x}{3}}[u_v(x,Q_0^2)-d_v(x,Q_0^2)]$ are evaluated
for a comparison with the available experimental data.
In sec-V; we attempt to obtain the gluon and the sea quark distributions
$G(x,Q_0^2)$ and $q_s(x,Q_0^2)$ respectively by dynamically
generating them from the well known leading order renormalization
group equation [21,22] with the valence distributions as the
inputs. Then we evaluate the momentum fraction carried by the
quark sea, the gluons and the valence quarks at $Q_0^2=15GeV^2$
leading to the saturation of the momentum sum-rule. Finally 
to realize the complete structure functions $F_2^{p,n}(x,Q_0^2)$
and their difference $[F_2^p(x,Q_0^2)-F_2^n(x,Q_0^2)]$ taking 
into account appropriate sea contributions together with
the corresponding valence parts; we consider some specific
prescriptions for the flavor decomposition of the sea. The results 
are then compared with the available experimental data.At the end;
 sec-VI provides a brief summary and conclusion.
   
\section{MODEL FRAMEWORK}

In a parton model study of deep inelastic scattering (DIS) 
of electrons off the nucleon; which is pictured as three
valence quarks embedded in sea of virtual quark antiquark
pairs and gluons; the partons within the nucleon are treated 
as approximately free because of the asymptotic freedom
property of QCD-interaction and light cone dominance of 
DIS. But from the point of view of a phenomenological 
quark model to start with; it may be quite justified 
to consider the nucleon as consisting only of three 
valence quarks, which eventhough might be dressed by the
sea-quarks and the gluon; can be taken as the only resolvable 
individual units with no further discrenible internal 
structure at the hadronic scale of low $Q^2={\mu}^2$.
The gluon and the sea-quark contents at $Q^2\gg {\mu}^2$;
can be realized through dynamic generation via gluon 
bremsstrahlung and quark pair creation in the frame-work
of QCD. The valence quarks constituting the nucleon 
at the model scale; being bound by the confining interaction 
within the hadronic volume; are not really free to be 
in any definite momentum states. However in order to 
establish a link with the parton model picture of DIS; one
can argue in principle that the bound valence quarks in a
nucleon; during the virtual compton scattering envisaged
in the description of DIS; can be sensed by the interacting
virtual photon in various momentum states with certain 
probabilities appropriate to their bound state 
energy-eigen modes. These momentum probability amplitudes 
can be realized from the fourier projections of their energy
eigen modes. In that case the nucleon at the low resolution
scale can be thought of as a bundle of free valence quarks 
in SU(6) spin flavor configurations with some appropriate
momentum distribution satisfying in some heuristic manner 
the energy momentum conservation. Then one can analyse 
the deep inelastic scattering in terms of free valence quarks
interacting with the virtual photon at definite momentum states 
with specific momentum probabilities, which can enable one 
to establish a link between the low energy description of DIS
with the parton model interpretation at high energy.\\
\par In view of our above motivation; we prefer to represent 
the nucleon in its ground state with a definite momentum 
${\bf P}$ and spin projection ${\bf S}$, to a first approximation;
by a normalized momentum wave packet of free valence quarks 
in the form; 
\begin{eqnarray}
\mid P,S> ={1\over {\sqrt {{\cal N}(P)}}}
\int {\prod\limits_{i=1}^{3}}{d^3{\bf k_{i}}\over {\sqrt {2E_{k_{i}}}}}
G_N({\bf k_{1}},{\bf k_{2}},{\bf k_{3}}) \nonumber \\
\delta^4(k_1+k_2+k_3-P)\mid T({\bf k_{1}},{\bf k_{2}},{\bf k_{3}};S)>
\end{eqnarray}
Here; $\mid T({\bf k}_1,{\bf k}_2,{\bf k}_3; S)>$ provides the SU(6) spin
flavor configuration of the valence quarks
in definite momentum states expressed as;
\begin{eqnarray}
\mid T({\bf k_{1}},{\bf k_{2}},{\bf k_{3}}; S)>&=&\sum\limits_{1\to 2,3}
{\cal Z}_{q_1,q_2,q_3}^N(\{ {\lambda}_i\} \in S)\nonumber \\
&\times & a_{q_1}^{\dagger}({\bf k}_1,{\lambda}_1)
a_{q_2}^{\dagger}({\bf k}_2,{\lambda}_2)
a_{q_3}^{\dagger}({\bf k}_3,{\lambda}_3) \mid 0>
\end{eqnarray}
we must mention here that ;
${\cal Z}_{q_1,q_2,q_3}^N(\{ {\lambda}_i\} \in S)$
denotes the usual spin flavor co-efficients
and ${\hat{a}}_{q}({\bf k},{\lambda}),
{\hat{a}}_{q}^{\dagger}({\bf k},{\lambda})$ are
respectively the free quark annihilation
and creation operators with definite momentum ${\bf k}$ and spin
projection `${\lambda}$'; which obey the usual anti
commutation relations. Finally $G_N({\bf k_{1}},{\bf k_{2}},{\bf k_{3}})$
represents the momentum profile function of the three quarks
which is subjected to the constraint of energy momentum 
conservation provided through the delta function in an adhoc manner.
If we consider $G_q({\bf k},{\lambda}^{'})$ as the momentum 
probability amplitude of the bound valence quark `q' in its lowest 
energy eigenmode ${\Phi}_{q{\lambda}}^{+}(r)$; to be found in a 
free state of definite momentum ${\bf k}$ and spin projection
${\lambda}^{'}$; then 
\begin{eqnarray}
G_q({\bf p},{\lambda}^{'})&=&{\frac{1}{{(2\pi)}^{\frac{3}{2}}}}
{\frac{u_q^{\dagger}({\bf p},{\lambda}^{'})}{\sqrt{2E_p}}}
\int d^3{\bf r}\;{\Phi}_{q{\lambda}}^{+}({\bf r})\;
exp{(-i{\bf p}{\bf r})}\nonumber\\
&=& G_q({\bf p}){\delta}_{{\lambda}{\lambda}^{'}}
\end{eqnarray}
where; $E_p={\sqrt{|{\bf p}|^2+m_q^2}}$ and $u_q({\bf p},{\lambda}^{'})$
is the usual free Dirac spinor. With reference to a specific 
phenomenological quark model such as the independent quark model 
with scalar vector harmonic potential [15], $G_q({\bf k},{\lambda}^{'})$
can be worked out in the form [16] as;
\begin{eqnarray}
G_q({\bf k},{\lambda}^{'})=G_q({\bf k}){\delta}_{{\lambda}{\lambda}^{'}}
\end{eqnarray}
when;
\begin{eqnarray}
G_q({\bf k})={i{N_q}r^2_{0q}\over {2\sqrt {2\pi}}}{(E_p+E_q)
\over {\lambda_q}} \Biggl[\frac{(E_p+m_q)}{E_p}\Biggr]^{1/2}
\exp {(-{\frac{r^2_{0q}|{\bf k}|^2}{2}})}
\end{eqnarray}
Here $E_q$ is the ground state binding energy of the bound quark 
in the potential field $ V(r)=(1/2)(1+{\gamma}^0)(ar^2+V_0)$
with $r_{0q}={(a{\lambda}_q)}^{-{\frac{1}{4}}}$; and 
\begin{equation}
N_q^2={\frac{8(E_q+m_q)}{\sqrt{\pi}r_{0q}(3E_q+m_q-V_0)}}
\end{equation}
Then the momentum profile function 
$G_N({\bf k_{1}},{\bf k_{2}},{\bf k_{3}})$  of the three quarks 
in the nucleon can be expressed in the product form as ;

\begin{eqnarray}
G_N({\bf k_{1}},{\bf k_{2}},{\bf k_{3}})=G({\bf k_{1}})
G({\bf k_{2}})G({\bf k_{3}})
\end{eqnarray}

It may however be noted that the momentum probability amplitude 
$G_q({\bf k})$ of individual quarks would be flavor independent 
in the non-strange sector; since the model adopted here assumes SU(2)
flavor symmetry.  Finally we have taken an overall 
normalization factor in Eq.(2.1) as $[{\cal N}(P)]^{-1/2}$;
which can be determined considering the co-variant normalization 
condition,
\begin{eqnarray}
<P,S \mid P^{'},S^{'}>={(2\pi)}^32E_N\delta^3({\bf P}-{\bf P^{'}})
\delta_{SS^{'}}
\end{eqnarray}
Using Eq.(2.1) in Eq.(2.8) and expressing the momentum 
probability distribution for quark $q_{i}$ as ${\rho}_{i}
({\bf k})=|G_{i}({\bf k})|^2 $; one can obtain,
\begin{equation}
|{\cal N}({\bf P})|={\frac{{\delta}(0)}{16{\pi}^3E_N}}
{\tilde{I}}_N
\end{equation}
and;
\begin{equation}
{\tilde{I}}_N=\int\;\prod\limits_{i=1}^{3}{\frac{d^3{\bf k}_i}
{2E_{k_i}}}{\rho}({\bf k}_i){\delta}^4(k_1+k_2+k_3-P)
\end{equation}
The integral in Eq.(2.10) can be evaluated in a quark mass
limit $m_q{\longrightarrow}0$ for the nucleon at rest.
For doing this; we express the energy-delta function term appearing in
the expression as;
\begin{equation}
{\delta}(|{\bf k}_1|+|{\bf k}_2|+|({\bf k}_1+{\bf k}_2)|-M)=
{\frac{(M-|{\bf k}_1|-|{\bf k}_2|)}{|{\bf k}_1||{\bf k}_2|}}
{\delta}(z-{\bar z})
\end{equation}
Here $z=cos{\theta}_{k_2}$; which sets the limits of integrations
for $|{\bf k}_2|$ as ;
\begin{equation}
({\frac{M}{2}}-|{\bf k}_1|)\;{\leq}\;|{\bf k}_2|\;{\leq}\;{\frac{M}{2}}
\end{equation}
Then defining;
\begin{eqnarray}
{\tilde {\rho}}_{ij}(|{\bf k}|) &=& \int^{M\over 2}_{
{M\over 2}-|{\bf k}|}d|{\bf k}^{'}|
{\rho}_{i}(|{\bf k}^{'}|) {\rho}_{j}(M-|{\bf k}|-|{\bf k}^{'}|)\nonumber\\
I_N &=& \int_{0}^{\infty} d|{\bf k}| {\rho}_1(|{\bf k}|)
{\tilde {\rho}}_{23}(|{\bf k}|)
\end{eqnarray}
so that; the normalization constant for the nucleon state
corresponding to its rest frame can be found as;
\begin{eqnarray}
{\cal N}({\bf P}=0)={\frac{\delta(0)}{16\pi M}}I_N
\end{eqnarray}
The integrals for ${\tilde{\rho}}_{ij}(|{\bf k}|)$ and $I_N$ can 
either be evaluated analytically or numerically.

\section{STRUCTURE FUNCTIONS IN THE MODEL}

The hadronic tensor describing the deep-inelastic electron-nucleon 
scattering; which is expressed as the Fourier transform
of single nucleon matrix element of the commutator of two 
electromagnetic currents in the form;
\begin{equation}
W_{\mu\nu}={1\over {4\pi}}\int d^4{\xi}\;e^{iq{\xi}}<P,S\mid\
[J_{\mu}({\xi}),J_{\nu}(0)]\mid\ P,S>,             
\end{equation}
can be analysed in the present model frame-work to derive the
nucleon structure functions.  In Eq.(3.1)
q is the virtual photon four-momentum and (P, S) are respectively the four-
momentum  and spin of the target nucleon, 
such that
\begin{equation}
P^{\mu}P_{\mu}=M^2,\;S^{\mu}S_{\mu}=-M^2 \; and \; \;  P^{\mu}S_{\mu}=0
\end{equation}
The coventional kinematic variables are usually defined  as 
$Q^2=-q^2>0$ and $x=Q^2/2{\nu}$;  when ${\nu}=P{\cdot}q$ and
$0{\leq}x{\leq}1$.  In the rest frame of the target nucleon; one
takes $P{\equiv}(M,0,0,0)$ and $q{\equiv}({\nu}/M,0,0,\sqrt 
{{\nu}^2/M^2+Q^2})$.
\par The hadronic tensor in Eq.(3.1) can be decomposed into
a symmetric part $W_{{\mu}{\nu}}^{(S)}$ and an antisymmetric part 
$W_{{\mu}{\nu}}^{(A)}$ respectively; when $W_{{\mu}{\nu}}^{(S)}$ 
defines the spin averaged structure functions $F_1(x,Q^2)$ and
$F_2(x,Q^2)$ through a co-variant expansion in terms of the
scalar functions $W_1(x,Q^2)$ and $W_2(x,Q^2)$ as ;
\begin{equation}
W_{\mu\nu}^{(S)}={\Bigl[}-g_{\mu\nu}+{\frac{q_{\mu}q_{\nu}}{q^2}}{\Bigl]}
W_1(x,Q^2)+{\Bigl[}(P_{\mu}-q_{\mu}{\frac{P\cdot q}{q^2}})
(P_{\nu}-q_{\nu}\frac{P\cdot q}{q^2}){\Bigl]}\frac{W_2(x,Q^2)}{M^2}.
\end{equation}
The deep-inelastic unpolarized structure functions $F_1(x,Q^2)$
and $F_2(x,Q^2)$ which become the scaling functions of the
Bjorken variable $x$ in the Bjorken limit ($Q^2\longrightarrow\infty$,
and ${\nu}\longrightarrow\infty$; with $x$ fixed) are defined as 
$F_1(x,Q^2){\equiv}W_1(x,Q^2)$ and $F_2(x,Q^2){\equiv}
{\nu}W_2(x,Q^2)/M^2$.  It is well known that while  
$F_2(x,Q^2)$ provides the contributions
of the transverse virtual photons; a combination such as
$W_L(x,Q^2)=[F_2(x,Q^2)/2x-F_1(x,Q^2)]$ owes it to the
longitudinal virtual photons.  It can be shown that 
$W_L(x,Q^2)={\frac{2M^2x}{\nu}}W_{00}^S$; so that with $W_{00}^S$
as finite in the Bjorken limit; $W_L\longrightarrow 0$ satisfies
there-by the  so called Callen Gross relation 
\begin{equation}
F_2(x,Q^2)=2xF_1(x,Q^2).
\end{equation}
Now for a model derivation of the structure functions
one can start with Eq. (3.1) with a static no gluon approximation
for the target nucleon considered at rest with the nucleon state 
${\mid P,S>}$ represented as a momentum wave-packet of the
constituent valence quarks as given in Eq.(2.9). However it
is convenient to recast Eq.(3.1) into a more suitable form [5] as
\begin{eqnarray}
W_{\mu\nu}(q,S)={1\over {32{\pi}^4{\delta}^3(0)}} \int_{-\infty}^{+\infty} dt
e^{iq_0t} \int d^3{\bf r} \int d^3{\bf r}^{'} e^{-i{\bf q}\cdot ({\bf r}
-{\bf r}^{'})} \nonumber \\
\times <P,S \mid [J_{\mu}({\bf r},t),J_{\nu}({\bf r}^{'},0)] 
\mid P,S>.
\end{eqnarray}
The electromagnetic current of the target nucleon is taken here
in the form $J_{\mu}({\xi})=\sum\limits_{q}e_{q}{\bar{\psi}}_{q}({\xi})
{\gamma}_{{\mu}}{\psi}_{q}({\xi})$; where $e_q$ is the electric 
charge of the valence quark of flavor $q$ inside the  nucleon.
The quark field operators  ${\psi}_q({\xi})$ is expressed here
appropriately as the free field expansion;
\begin{equation}
{\psi}_q({\xi})={1\over {{(2\pi)}^{\frac{3}{2}}}}\sum\limits_{\pm\lambda}
\int\; {\frac{d^3 {\bf p}}{\sqrt{2E_p}}}[{\hat{a}}_q({\bf p},{\lambda})
u({\bf p},{\lambda}) e^{-ip{\xi}}+{\hat{\tilde{a}}}_q^{\dagger}
({\bf p},{\lambda})v({\bf p},{\lambda}) e^{ip{\xi}}]
\end{equation}
where the free Dirac spinors for the valence quarks taken in
the zero mass limit as;
\begin{eqnarray}
u({\bf p},{\lambda})={\sqrt{E_p}} \pmatrix{
1 \cr
{{\vec {\sigma}}\cdot {\vec {p}}} /E_p \cr} {\chi}_{\lambda}  \nonumber\\
v({\bf p},{\lambda})={\sqrt{E_p}} \pmatrix{
{{\vec {\sigma}}\cdot {\vec {p}}} /E_p \cr 1 \cr} {\tilde{{\chi}_{\lambda}}}  
\end{eqnarray}
Now expanding the curent commutator in Eq. (3.5) and taking
the free quark propagator appearing in the expansion 
under  impulse approximation written in the zero mass limit as;

\begin{equation}
{\lim \limits_{m\rightarrow 0}} S_D(x) =
{1\over {(2\pi)^3}} \int d^4k {\rlap/{k}} {\epsilon}(k_0)
{\delta}(k^2) e^{\pm ikx},      
\end{equation}

where ${\epsilon}(k_0)=sign(k_0)$; the symmetric part of the
hadronic tensor $W_{{\mu}{\nu}}^{(S)}$ can be obtained as;

\begin{equation}
W_{\mu\nu}^{(S)}=[g_{\mu\lambda}g_{\nu\sigma}
+g_{\mu\sigma}g_{\nu\lambda}
-g_{\mu\nu}g_{\lambda\sigma}]T^{{\lambda}{\sigma}}.   
\end{equation}
when
\begin{eqnarray}
T^{\lambda\sigma}={[32{\pi}^4{\delta}^3(0)]}^{-1} \sum\limits_{q}
e_q^2 \int {\frac{d^4k}{{(2\pi)}^3}} k^{\lambda} {\epsilon}(k_0)
{\delta}(k^2) \int_{-\infty}^{+\infty} dt e^{i(q_0+k_0)t} \nonumber \\
\times \int d^3{\bf r} d^3{\bf r}^{'} e^{-i({\bf q}+
{\bf k})\cdot ({\bf r}-
{\bf r}^{'})} < {\Lambda}^{\sigma} >,  
\end{eqnarray}
and 
\begin{eqnarray}
< {\Lambda}^{\sigma} >=< P,S \mid [{\bar {\psi}}_{q}({\bf r},t)
{\gamma}^{\sigma} {\psi}_{q}({\bf r}^{'},0) 
- {\bar {\psi}}_{q}({\bf r}^{'},0) 
{\gamma}^{\sigma}{\psi}_q({\bf r},t)]\mid P,S>. 
\end{eqnarray}
Since it is evident from Eq.(3.3) that $F_1(x,Q^2)\equiv W_1(x,Q^2)$ is the 
co-efficient of $(-g_{\mu\nu})$ in the covariant expansion of
$W_{\mu\nu}^{S}$; Eq.(3.9) in the same token  
can yield;
\begin{equation}
F_1(x,Q^2)=g_{{\lambda}{\sigma}}T^{{\lambda}{\sigma}} 
\end{equation}
Thus we find; 
\begin{eqnarray}
F_1(x,q^2)={[32{\pi}^4{\delta}^3(0)]}^{-1} 
\int {\frac{d^4k}{{(2\pi)}^3}} {\epsilon}(k_0)
{\delta}(k^2) \int dt e^{i(q_0+k_0)t} \nonumber \\
\times \int d^3{\bf r} d^3{\bf r}^{'} e^{-i({\bf q}+
{\bf k})\cdot ({\bf r} - {\bf r}^{'})} < {\Gamma} >,
\end{eqnarray}
where;
\begin{eqnarray}
<\Gamma>=<P,S \mid\sum\limits_{q}e_q^2[{\bar {\Psi}}_q({\bf r},t){\rlap/{k}}
{\Psi}_q({\bf r^{'}},0)-{\bar {\Psi}}_q({\bf r^{'}},0){\rlap/{k}}
{\Psi}_q({\bf r},t)] \mid P,S>
\end{eqnarray}
Now substituting the nucleon state $ \mid P,S> $ as in Eq.(2.1),(2.2), 
along with the free field expansions of the field operators as in 
Eq.(3.6),(3.7), we can realize after some necessary algebra 
\begin{equation}
F_1(x,Q^2)=[f_{+}(x,Q^2) - f_{-}(x,Q^2)]
\end{equation}
where;
\begin{eqnarray}
f_{\pm}(x,Q^2)={\frac{{\delta}(0) <e_q^2>}{16{\pi}^3{\cal N}(0)}}
\int dk_0k_0{\epsilon}(k_0) \int d^3{\bf k} {\frac{{\delta}
(|{\bf k}|+k_0)}{2|{\bf k}|}} \nonumber\\
\times \int {\frac{d^3{\bf k}_1}{2E_{k_1}}}{\rho}(|{\bf k}_1|)
{\Bigl[}1+{\frac{{\hat{{\bf k}}}{\cdot}{\bf k}_1}{E_{k_1}}}
+i{\frac{({\hat{{\bf k}}}{\times}
{\bf k}_1)_z}{E_{k_1}}}{\Bigl]}{\delta}^3({\bf k}+{\bf q}{\pm}{\bf k}_1)
{\delta}(q_0+k_0{\pm}E_{k_1}) \nonumber\\
\times \int {\frac{d^3{\bf k}_2d^3{\bf k}_3}{4E_{k_2}E_{k_3}}}
{\rho}(|{\bf k}_2|){\rho}(|{\bf k}_3|){\delta}^3({\bf k}_1
+{\bf k}_2+{\bf k}_3){\delta}(E_{k_1}+E_{k_2}+E_{k_3}-M)
\end{eqnarray}
It is to be noted here that, with the SU(2) flavor symmetry assumed 
in the present model; the spin flavor sum of the square of the quark
charges of each flavor weighted by the respective probability
$|{\cal Z}_{q_i}^N|^2$ corresponding to
its SU(6) configuration denoted here as $<e_q^2>$;
gets decoupled from the rest of the integrals after simplification.
Then one can independently evaluate $<e_q^2>$ for the spin up
proton target as 1 and the same for the neutron target as 2/3.

\par In order to be able to perform the $k_0$-integration 
first amongst the nested integrals, we first make a reasonable 
approximation to extract ${\delta}(k_0+q_0{\pm}E_{k_1})$
from within the ${\bf k}_1$-integration as ${\delta}
(k_0+q_0{\pm}{\bar E})$ with $E_{k_1}{\simeq}{\bar E}$
corresponding to the peak position of the momentum distribution
${\rho}({\bf k}_1)$ under the expressions for $f_{\pm}(x,Q^2)$.
It would then imply $k_0=-(q_0{\pm}{\bar E})$; which are
always negative in the Bjorken limit for $f_{+}$ \&
$f_{-}$ respectively. Then putting ${\bf K}={\bf q}+{\bf k}$
so that $|{\bf K}|=K\geq K_m=(|{\bf q}|-|{\bf k}|)$;
where $K_m$ can reasonably be assumed to be much less
than $(q_0,|{\bf q}|\; \& \; |{\bf k}|)$ in the Bjorken limit.
Now doing the $k_0$-integration; we can write for the proton;
\begin{eqnarray}
f_{\pm}^p(x,Q^2)={\frac{{\delta}(0)}{16{\pi}^3{\cal N}(0)}}
\int\;d{\phi}_k d(cos{\theta}_{k})\int_{0}^{\infty}
{\frac{d|{\bf k}|\;|{\bf k}|^2}{2}}
{\delta}(q_0-|{\bf k}|\pm{\bar E})\nonumber\\
\times \int {\frac{d^3{\bf k}_1}{2E_{k_1}}}{\rho}(|{\bf k}_1|)
{\Bigl[}1+{\frac{{\hat{{\bf k}}}{\cdot}{\bf k}_1}{E_{k_1}}}
+i{\frac{({\hat{{\bf k}}}{\times}
{\bf k}_1)_z}{E_{k_1}}}{\Bigl]}{\delta}^3({\bf K}{\pm}{\bf k}_1)\nonumber\\
\times \int {\frac{d^3{\bf k}_2d^3{\bf k}_3}{4E_{k_2}E_{k_3}}}
{\rho}(|{\bf k}_2|){\rho}(|{\bf k}_3|){\delta}^3({\bf k}_1
+{\bf k}_2+{\bf k}_3){\delta}(E_{k_1}+E_{k_2}+E_{k_3}-M)
\end{eqnarray}
The delta function ${\delta}(q_0-|{\bf k}|{\pm}{\bar E})$ 
in Eq.(3.17) sets the value of $|{\bf k}|=k$ as 
$k_{\pm}=q_0{\pm}{\bar E}$ and ${\delta}^3({\bf K}{\pm}{\bf k}_1)$
sets the struck quark momentum as ${\bf k}_1={\mp}{\bf K}$.
This now leads to certain kinematic relations relevant in further 
simplifying the expression $f_{\pm}^p(x,Q^2)$ in the Bjorken 
limit, which are as follows;
\begin{eqnarray}
K_m\equiv{\bar K}_{\pm}(x)&=&|({\bar E}{\mp}Mx)|, \nonumber \\
\cos {\theta}_K &{\simeq}& (Mx {\mp}{\bar E})/K, \nonumber \\
\cos {\theta}_K \cos {\theta}_k &{\simeq}&
-(Mx {\mp}{\bar E})/K,  \nonumber \\
d({\cos {\theta}_k})k_{\pm}^2 &{\simeq}&KdK.
\end{eqnarray}
Now using these kinematic relations to-gether with the same
procedure as described in Eq.(2.11) and (2.12) and finally 
substituting ${\cal N}(0)$ as in Eq.(2.13) to (2.14) after the 
necessary simplifications; we get
\begin{equation}
f_{\pm}^p(x,{\mu}^2)={M\over {4I_N}} \int^{\infty}_{{\bar K}_{\pm}(x)}
{dK\over {K^2}}\rho (K){\tilde \rho}(K)[K-K_{\pm}(x)]
\end{equation}
where we have used $K_{\pm}(x)=({\bar E}\mp Mx)$. It may be noted here
that ${\tilde {\rho}}(K)$ represents the effects of the spectator
quarks.
\par Thus using Eq.(3.19) in (3.15); we can obtain the structure
function $F_1^p(x,Q^2)$ for the proton at its bound state scale.
Similar calculation can lead to $F_1^n(x,Q^2)$ for the neutron; 
which would be ${\frac{2}{3}}F_1^p(x,Q^2)$ in the present model
with SU(2) flavor symmetry. Since as usual it can be shown here that 
$W_{00}^{S}(x,Q^2)$ is finite [24] in 
the Bjorken limit which would lead to 
$W_L{\longrightarrow}0$ satisfying the Callen-Gross relation 
from which $F_2^{p,n}(x,Q^2)$ can also be realized using the 
expressions derived for $F_1^{p,n}(x,Q^2)$.

\section{VALENCE QUARK DISTRIBUTION FUNCTIONS}

In a parton picture, if we define the quark
parton distribution functions in the
(u,d) flavor sector inside the nucleon in the usual manner as 
a combination of valence and sea components, such as,
$u(x,Q^2)=u_v(x,Q^2)+u_s(x,Q^2)$ and 
$d(x,Q^2)=d_v(x,Q^2)+d_s(x,Q^2)$ with the corresponding 
antiparton distributions defined accordingly; then 

\begin{eqnarray}
F_1^p(x,Q^2)&=&1/18[\{4u(x,Q^2)+d(x,Q^2)\}+\{4{\bar u}(x,Q^2)
+{\bar d}(x,Q^2)\}] \nonumber\\
F_1^n(x,Q^2)&=&1/18[\{4d(x,Q^2)+u(x,Q^2)\}+\{4{\bar d}(x,Q^2)
+{\bar u}(x,Q^2)\}]
\end{eqnarray}

\par Now comparing expressions in Eq. (4.1) with Eq. (3.15) and
 attributing as usual for such models ;
 the negative part of the distributions in Eq.(3.15)
 to the anti-partons in Eq.(4.1); effective
parton distributions can be identified [5] as;

\begin{eqnarray}
u(x,Q^2)&=&2d(x,Q^2)=4f_{+}^P(x,Q^2) \nonumber\\
{\bar u}(x,Q^2)&=&2{\bar d}(x,Q^2)=-4f_{-}^P(x,Q^2)
\end{eqnarray}

\par It is to be noted here that the negative antiparton 
distributions so obtained at the model scale calculation,
can be treated only as a model artifact which infact is
encountered in all such constituent quark models [5].
This spurious contribution needs to be appropriately
eliminated in extracting the valence quark distribution
correctly from the effective parton distributions 
in Eq.(4.2).  Thus keeping in
mind that ${\bar u}_v(x,Q^2)=0={\bar d}_v(x,Q^2)$ as per 
our initial assumption and considering the spurious parton 
and anti-parton sea to be 
symmetric(i.e $u_s(x)={\bar u}_s(x)={\bar u}(x)$ and
$d_s(x)={\bar d}_s(x)={\bar d}(x)$ etc ); we get
the appropriate valence distributions as

\begin{equation}
u_v(x,Q^2)=2d_v(x,Q^2)=4[f_{+}^p(x,Q^2)+f_{-}^p(x,Q^2)].
\end{equation}

Thus the valence quark distribution functions $u_v(x,Q^2)$
and $d_v(x,Q^2)$ can be extracted at a model scale of low
$Q^2={\mu}^2$ in terms of analytically obtained 
expressions $f_{\pm}^P(x,Q^2)$ as functions of the Bjorken
variable $x$; which can be evaluated by taking the model
parameters $(a,V_0)$ and other relevant model quantities
such as $(m_q,E_q,r_{0q}\; etc)$ described in sec-II
as per their values found in its 
earlier applications in Ref [15,16] such as;

\begin{eqnarray}
(a,V_0)&=&(0.017166GeV^3,-0.1375GeV),\nonumber \\
(m_q=m_u=m_d,\;E_q,\;r_{0q})&=&(0.01GeV,0.45129 GeV,3.35227 GeV^{-1})
\end{eqnarray}
However; in view of such a current quark mass limit 
adopted in the model applications earlier; we believe
in the justification of making all our calculations
meant for the ultimate Bjorken limit with $m_q{\longrightarrow}0$ 
on the grounds of derivational simplicity. In that case the 
corresponding model quantities $E_q$ and $r_{0q}$, 
relevant for our calculations, are not much different 
from those given in Eq.(4.4); since their values now would be;

\begin{equation}
(E_q,r_{0q})=(0.4490 GeV,3.37489 GeV^{-1})
\end{equation}

with the same potential parameters $(a,V_0)$ as in Eq.(4.4).

\par we take here the actual physical mass of the proton
$M=0.940 GeV$ and ${\bar E}{\simeq}0.18 GeV$ 
corresponding approximately to
the peak position of the momentum distribution ${\rho}(|{\bf k}_1|)$.
The distributions $xu_v(x,Q^2)$ and $xd_v(x,Q^2)$ are evaluated 
numerically as functions of $x$ which are presented in Fig.1
and Fig.2 respectively showing correct support. It is found that 
these distribution functions for the valence quarks satisfy
the normalization requirement as;

\begin{equation}
\int_0^1\;dx \pmatrix{u_v(x,Q^2) \cr
d_v(x,Q^2) \cr}=\pmatrix{1.999\cr 0.999\cr}\simeq\pmatrix{
2\cr 1\cr}
\end{equation}
while the total momentum carried by the valence quarks at this low 
reference scale comes out as;

\begin{equation}
\int_0^1\;x[u_v(x,Q^2)+d_v(x,Q^2)]dx=0.994\simeq 1
\end{equation}
Thus with a close consistency in the requirements of 
normalization and momentum saturation at the model scale;
it can be justfied to use these valence distributions 
as appropriate model scale inputs for QCD-evolution to
higher $Q^2$. Realizing the valence distributions at
experimentally relevant higher $Q^2$-region 
through QCD evolution; one
can further evaluate the valence parts of the structure 
functions such as $F_2^p(x,Q^2)]_v={\frac{1}{2}}xu_v(x,Q^2)$
and $[F_2^n(x,Q^2)]_v={\frac{1}{3}}xu_v(x,Q^2)$ as well
as the valence part of the combination $[F_2^p(x,Q^2)-F_2^n(x,Q^2)]_v
={\frac{1}{6}}xu_v(x,Q^2)$.

\par However the model scale of low $Q^2={\mu}^2$ is neither explicit 
in the derived expressions for the structure functions nor
in the valence distributions $u_v(x,Q^2)$ and $d_v(x,Q^2)$.
Therefore we need to first fix the model scale $Q^2={\mu}^2$;
with the help of the renormalization group
equation [13],as per which;
\begin{equation}
 {\mu}^2={\Lambda}_{QCD}^2e^L; 
\end{equation}
where;
$ L=[V^{n=2}(Q_0^2)/V^{n=2}({\mu}^2)]^{1/a_{NS}^{n=2}}
ln({\frac{Q_0^2}{{\Lambda}_{QCD}^2}})  $ and
$V^{n=2}(Q^2)=\int_0^1dx x [u_v(x,Q^2)+d_v(x,Q^2)]$ as
the momentum carried by the valence quarks at $Q^2$.  Now 
taking the experimental reference scale $Q_0^2=15GeV^2$
for which $V^{n=2}(15GeV^2){\simeq}0.4$ [2,8] and 
$V^{n=2}({\mu}^2){\simeq}1 $ as in Eq.(4.7)
together with ${\Lambda}_{QCD}=0.232GeV$ and $a_{NS}^{n=2}=32/81$
for 3-active flavors; one can obtain ${\mu}^2=0.1GeV^2$.  
If one believes that the perturbation theory still makes 
sense down to this model scale for which the relevant 
perturbative expansion parameter ${\alpha}_s({\mu}^2)/2{\pi}$
is less than one ($\simeq0.358$), one can evolve
the valence distributions $u_v(x,{\mu}^2)=2d_v(x,{\mu}^2)$
to higher $Q_0^2$, where experimental data are available.
In fact one does not have much choice here, because taking any 
higher model scale on adhoc basis would require a non-zero
initial input sea quark and gluon constituents for which 
one does not have any dynamical information at such scale
and hence it would complicate the picture.  Therefore when
${\alpha}_s({\mu}^2)/2{\pi}$ is well within the limit to 
justify the applicability of perturbative QCD at the 
leading order and further since non-singlet evolution is
believed to converge very fast [23], to remain
stable even for small values of $Q^2/{\Lambda}^2_{QCD}$; one
may think of a reliable interpolation between the low model
scale of $Q^2={\mu}^2<0.1GeV^2$ and the experimentally
relevant higher $Q^2>>{\mu}^2$, if one does not insist
upon quantitative precision.  With such justification
and belief many authors in past have used the choice of low 
$Q^2={\mu}^2$(for example; ${\mu}^2=0.063GeV^2$ [9],
 $0.068GeV^2, 0.09GeV^2$ [10] and
$0.06GeV^2$ [12]) as their static point
for evolution.  Infact the choice of low $Q^2={\mu}^2=0.1GeV^2$
in such models is linked with the initial 
sea and gluon distributions being taken approximately zero at the
model scale.  Following such arguments; we choose
to evolve the valence distributions by the 
standard convolution tecnique based on nonsinglet evolution 
equations in leading order [3,23] from the static point of 
${\mu}^2=0.1GeV^2$ to $Q_0^2=15GeV^2$ for a comparison with
the experimental data.  Our results for $xu_v(x,Q_0^2)$ and 
$xd_v(x,Q_0^2)$ at $Q_0^2=15 GeV^2$ are provided in Fig. 1 and
Fig. 2 respectively along with the experimental data, which on
comparison shows satisfactory agreement over the
entire range $0\leq x\leq 1$. The valence components of the
structure functions such as $[F_2^p(x,Q_0^2)]_v$ and
$[F_2^n(x,Q_0^2)]_v$ together with the valence part of the
combination $[F_2^p(x,Q_0^2)-F_2^n(x,Q_0^2)]_v$ 
calculated at $Q_0^2=15GeV^2$,
are also compared with the respective experimental data in
Fig. [3,4 $\&$,5] respectively.  We find that the agreement 
with the data in all these cases are reasonably  
better in the region  $x>0.2$.  This is because in the 
small $x$ region; the sea contributions to the structure functions 
not included in the calculation so far; are
significant enough to generate the appreciable
departures from the data as observed here. 

\par Therefore for a complete description of the nucleon 
structure functions and hence the parton distributions 
in the nucleon; the valence contributions discussed above 
need to be supplemented by the expected gluon and sea-quark
contributions at high energies.

\section{GLUON AND SEA QUARK DISTRIBUTIONS}

The gluon and the sea-quark distributions at high energy inside
the nucleon can be generated purely radiatively with appropriate 
input of the valence distributions, using the well known leading
order renormalization group(R.G)-equations [22,23].
Considering that at higher energy,heavier flavors may be excited
above each flavor threshold, we define the total sea
quark distribution here upto three flavors as;
\begin{equation}
q_s(x,Q^2)=2[u_s(x,Q^2)+d_s(x,Q^2)+s_s(x,Q^2)]
\end{equation}
and the gluon distribution by $G_(x,Q^2)$.
Their moments $q_s^n(Q^2)$ and $G^n(Q^2)$ respectively can be
obtained in terms of the corresponding moment $V^n(Q^2)$ of the
input valence distributions $V(x,Q^2)=[u_v(x,Q^2)+d_v(x,Q^2)]$
according to the RG-equations such as:
\begin{eqnarray}
G^n(Q^2)=[{\frac{{\alpha}^n(1-{\alpha}^n)}{{\beta}^n}}L_0^{a_{NS}^n}
\{L_0^{-a_{-}^n}-L_0^{-a_{+}^n}\}]V^n(Q^2),
\end{eqnarray}

\begin{eqnarray}
q_s^n(Q^2)=[L_0^{a_{NS}^n}\{{\alpha}^nL_0^{-a_{-}^n}+
(1-{\alpha}^n)L_0^{-a_{+}^n}-L_0^{-a_{NS}^n}\}]V^n(Q^2),
\end{eqnarray}

where the n-th moments of the functions $A(x,Q^2){\equiv}
\{G(x,Q^2),q_s(x,Q^2),V(x,Q^2)\}$ are defined as

\begin{eqnarray}
A^n(Q^2)=\int_0^1dx x^{n-1} A(x,Q^2),
\end{eqnarray}
and the RG-exponents such as $\{{\alpha}^n,{\beta}^n,
a_{NS}^n,a_{\pm}^n \}$ in the conventional notations are derivable
for the n-th moment as per Ref [23].  Finally
$L_0={\frac{{\alpha}_s({\mu}^2)}{{\alpha}_s(Q_0^2)}}
={\frac{ln(Q_0^2/{\Lambda}_{QCD}^2)}{ln({\mu}^2/{\Lambda}_{QCD}^2)}}$
which can also be expressed here in terms of
the momentum carried by the valence quarks
at $Q_0^2$ on the basis of the  momentum saturation
by valence quarks at the model scale $Q^2={\mu}^2$ as;

\begin{equation}
L_0={\Bigl[}\int_0^1 dx x V(x,Q_0^2){\Bigl]}^{-1/a_{NS}^{n=2}};
\end{equation}

 With $a_{NS}^{n=2}=32/81$ for three active flavors
 considered here,the value of $L_o$ comes out as 
 $L_0\approx 9$. Then calculating
the appropriate RG-exponents as per Ref [23] for
$n=2,4,6,8$ (higher moments being significantly smaller
are not considered here) and the corresponding moments
$V^n(Q_0^2=15GeV^2)$ from the evolved valence distribution
$u_v(x,Q_0^2)=2d_v(x,Q_0^2)$ at $Q_0^2=15GeV^2$; we evaluate
the respective moments $G^n(Q_0^2)$ and $q_s^n(Q_0^2)$ from
Eqns (5.2) and (5.3).  Then gluon and 
sea-quark distributions can be extracted 
by a  matrix inversion technique with the help of
simple parametric expressions taken for $xG(x,Q_0^2)$ and
$xq_s(x,Q_0^2)$ as;

\begin{equation}
$$xG(x,Q_0^2)=[a_1x^2+a_2x+a_3+a_4/\sqrt{x}~~] $$
\end{equation}

\begin{equation}
$$xq_s(x,Q_0^2)=[b_1x^2+b_2x+b_3+b_4/\sqrt{x}~~] $$
\end{equation}
The moments calculated from these parametric expressions
would now provide a set of simultaneous equations for each set of
parameters $\{a_i\}$ and $\{b_i\}$ separately. Solving these
equations by matrix inversion method we arrive at the values
of these parameters as:
\begin{eqnarray}
\{a_i, i=1,2,3,4\}{\equiv}(-0.8659,\; 2.0447,\; -2.1086,\; 0.9223)\nonumber\\
\{b_i, i=1,2,3,4\}{\equiv}(-0.2229,\; 0.5093,\; -0.5007,\; 0.2123)
\end{eqnarray}

Thus we generate somewhat reasonable functional forms for
$xq_s(x,Q_0^2)$ and $xG(x,Q_0^2)$ at $Q_0^2=15GeV^2$ which
are provided in Fig. 6 and Fig. 7 respectively in comparison
with the available experimental data.  We find the qualitative
agreement with the data quite encouraging with almost
vanishing contributions in both cases beyond $x>0.5$.

\par  We find next the momentum distributions 
for different constituent partons at $Q_0^2=15GeV^2$ by
calculating the second moments of the distribution functions
$u_v(x,Q_0^2), d_v(x,Q_0^2), q_s(x,Q_0^2)$ and $G(x,Q_0^2)$
respectively so as to obtain them as;

\begin{eqnarray}
u_v(Q_0^2)&=&0.279,  (0.275 \pm 0.011)\nonumber\\
d_v(Q_0^2)&=&0.140,  (0.116 \pm 0.017)\nonumber\\
q_s(Q_0^2)&=&0.106,  (0.074 \pm 0.011)\nonumber \\
G(Q_0^2)  &=&0.475,  (0.535)
\end{eqnarray}

For a comparison; the experimental values are shown within the
brackets against the calculated values. We
find that the parton distributions realized at a
qualitative level in the model at $Q_0^2=15GeV^2$;
saturate the momentum sum-rule. Finally to evaluate
the complete structure functions
$F_2^{(p,n)}(x,Q_0^2)$ by supplementing
the respective valence components with the necessary
sea-contributions; we consider a flavor decomposition
of the net sea-quark distribution $q_s(x,Q_0^2)$. With an
old option of a complete symmetric sea in SU(3)-flavor
sector;
\begin{equation}
[F_2^p(x,Q_0^2)]_{sea}=[F_2^n(x,Q_0^2)]_{sea}={2\over 9}xq_s(x,Q_0^2)
\end{equation}
\par However it has been almost established experimentally
that the nucleon quark sea is flovor asymmetric both in SU(2)
as well as SU(3) sector. Experimental violation
of Gottfried sum rule [25] and more recent and precise
asymmetry measurements in the Drell-Yan process with
nucleon targets [26] have shown a
strong x-dependence of the ratio $[d_s(x,Q^2)/u_s(x,Q^2)]$
with $d_s(x,Q^2)>u_s(x,Q^2)$ for $x<0.2$ and
$d_s(x,Q^2)$ running closer to $u_s(x,Q^2)$ for
$x>0.2$; whereas around $x=0.18$; $d_s(x,Q^2){\simeq}2u_s(x,Q^2)$.
Neutrino charm production experiment by CCFR
collaboration [26] also provides evidence in favor of
the relative abundance of strange to non-strange sea quarks
in the nucleon measured by a factor
${\kappa}{\equiv}{\frac{2<xs_s>}{<[xu_s+xd_s]>}}=0.477{\pm}0.063$.
Therefore keeping these experimental facts in mind;
we make a reasonable choice for the flavor structure of the
sea quark distribution as defined in Eq. (5.1) by taking;
\begin{eqnarray}
d_s(x,Q_0^2)=2u_s(x,Q_0^2)  \nonumber \\
s_s(x,Q_0^2)={1\over 4}[u_s(x,Q_0^2)+d_s(x,Q_0^2)]
\end{eqnarray}

Then we find the sea contributions to the structure functions
$F_2^{(p,n)}(x,Q_0^2)$ as;
\begin{eqnarray}
{[F_2^p(x,Q_0^2)]}_{sea}={1\over 5}xq_s(x,Q_0^2)  \nonumber \\
{[F_2^n(x,Q_0^2)]}_{sea}={13\over 45}xq_s(x,Q_0^2)
\end{eqnarray}

Then the complete structure functions
$F_2^p(x,Q_0^2)$ and $F_2^n(x,Q_0^2)$ with the valence
and quark sea components taken together are calculated and shown
in Fig.3 and Fig.4 .We find that the overall
qualitative agreement is reasonable for the region
$x>0.04$. We have also shown in Fig.5; the structure function combination
$[F_2^p(x,Q_0^2)-F_2^n(x,Q_0^2)]$ by taking into account
the asymmetric sea contribution as in Eq. (5.12), which
provides a relatively better agreement with the
available experimental data taken over a $Q^2$-range [2].

\section{SUMMARY AND CONCLUSION }

Starting with a constituent  quark model of 
relativistic independent quarks in an effective scalar-vector
harmonic potential and representing the nucleon as a 
suitably constructed wave-packet of free valence quarks 
only of appropriate momentum probability amlpitudes 
corresponding to their respective bound-state eigen-modes;
we have been able to analytically derive the deep-inelastic
unpolarized structure function $F_1^p(x,Q^2)$ at the model
scale of low $Q^2={\mu}^2=0.1GeV^2$ with correct support.
The valence quark distributions $u_v(x,Q^2)$ and $d_v(x,Q^2)$
have been appropriately extracted taking the parton model
interpretation of $F_1^p(x,Q^2)$. The valence distributions
$u_v(x,Q^2),d_v(x,Q^2)$ satisfy the normalization requirement 
as well as the momentum sum-rule constraints; providing there-by
suitable low energy model inputs for QCD-evolution to 
experimentally relevant $Q^2=Q_0^2=15GeV^2$.The valence 
distributions in the form $xu_v(x,Q_0^2)$ and $xd_v(x,Q_0^2)$;
the valence components $[F_2^{p,n}(x,Q_0^2)]_v$ as well
as the valence part of the combination $[F_2^p(x,Q_0^2)-
F_2^n(x,Q_0^2)]$ are then realized through the QCD-evolution
at $Q_0^2=15GeV^2$; which compare reasonably well with the 
experimental data in the expected range of the Bjorken variable.

\par The gluon distribution $G(x,Q_0^2)$ and the total
sea-quark distribution $q_s(x,Q_0^2)$ are dynamically 
generated from the renormalization group equations 
taking the moments of the valence quark distributions at
$Q_0^2=15GeV^2$ as inputs. The results for $xG(x,Q_0^2)$
and $xq_s(x,Q_0^2)$ find good agreement with the experimental 
data. Calculation of the constituent parton momenta also
yields the momentum percentage in the valence-quark
sector as ${\approx}27.9\%$ and ${\approx}14.0\%$ for the `u' and `d' flavor
quarks respectively; whereas in the sea-quark and gluon sector
we find the same to be ${\approx}47.5\%$ and ${\approx}10.6\%$ respectively 
saturating there-by the expected momentum sum-rule.
Incorporating the sea-quark contributions to the valence part
of the structure functions; the complete unpolarized 
structure functions $F_2^p(x,Q_0^2),F_2^n(x,Q_0^2)$ and 
the combination $[F_2^p(x,Q_0^2)-F_2^n(x,Q_0^2)]$ are obtained 
in reasonable agreement  with the data in the region
$x>0.1$.
\par There are of course various finer features of the nucleon
structure functions to-gether with their behaviour near the region
$x=0$; which would be beyond the limit of this simplistic approach 
in the model to address. Nevertheless,within its limitations,
the model is found to provide a simple parameter free analysis 
of the deep-inelastic unpolarized structure functions of the 
nucleon leading to the realization of its constituent 
parton distributions at $Q_0^2=15GeV^2$ with an over-all
qualitative agreement with experimental data.

\acknowledgements

We are  thankful to the Institute of Physics,Bhubaneswar,
INDIA; for providing necessary library and computational facilities
for doing this work. One of us (Mr R.N.Mishra) would like to  thank
Dr P.C.Dash of the Dept.of Physics, Dhenkanal College,Dhenkanal
for his helpful gesture and co-operation  during
this work.

\begin{figure}[t]

\protect\centerline{\epsfxsize=5.5in \epsfysize=6.5in \epsfbox{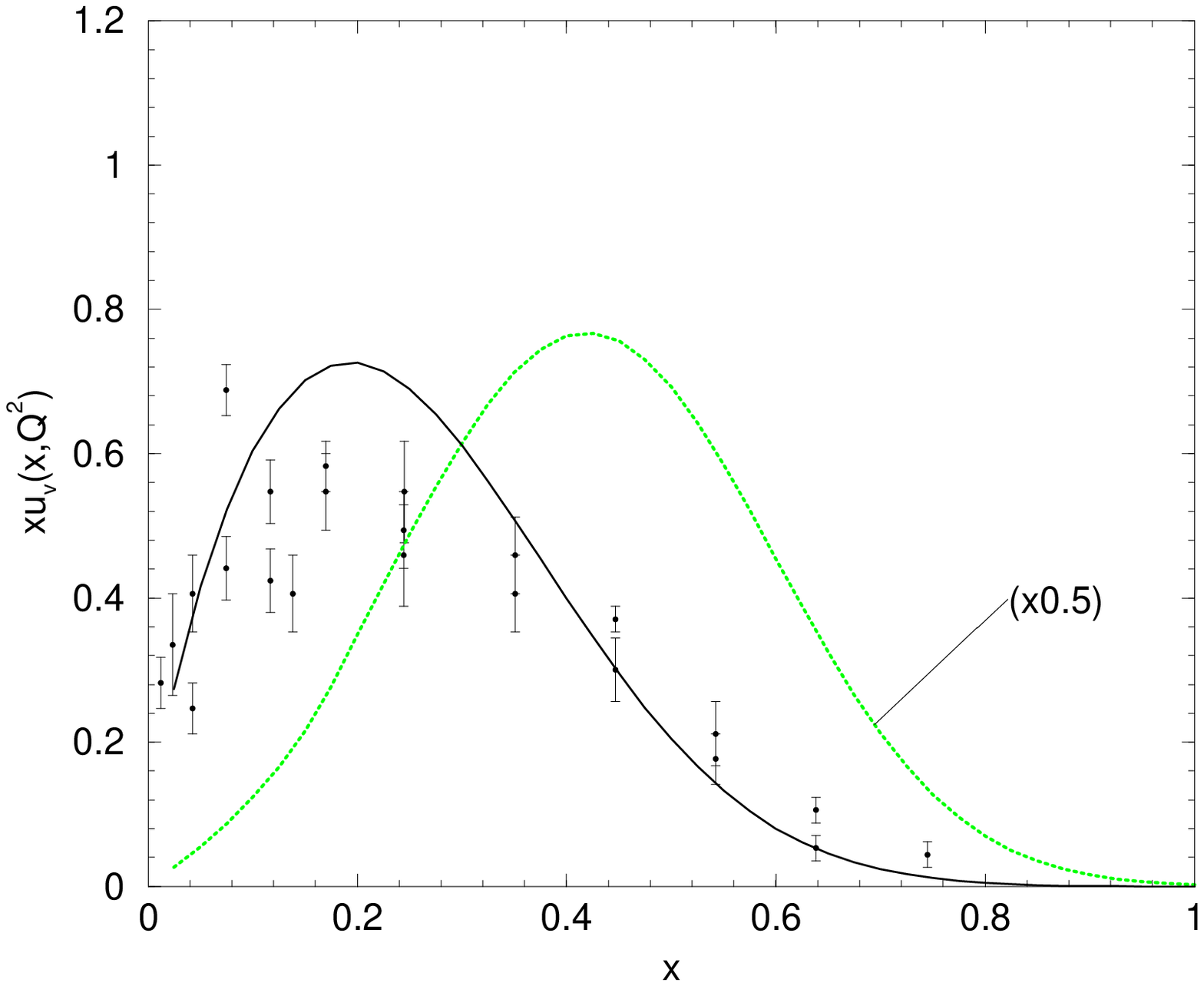}}

\caption{The calculated $xu_v(x,Q^2)$ at $Q^2={\mu}^2=0.1GeV^2$
(dotted line) and QCD evolved result at $Q_0^2=15GeV^2$ (solid line)
compared with the data taken from T.Sloan {\it et~al} in Ref.[2] }

\end{figure}

\begin{figure}[t]

\protect\centerline{\epsfxsize=5.5in \epsfysize=6.5in \epsfbox{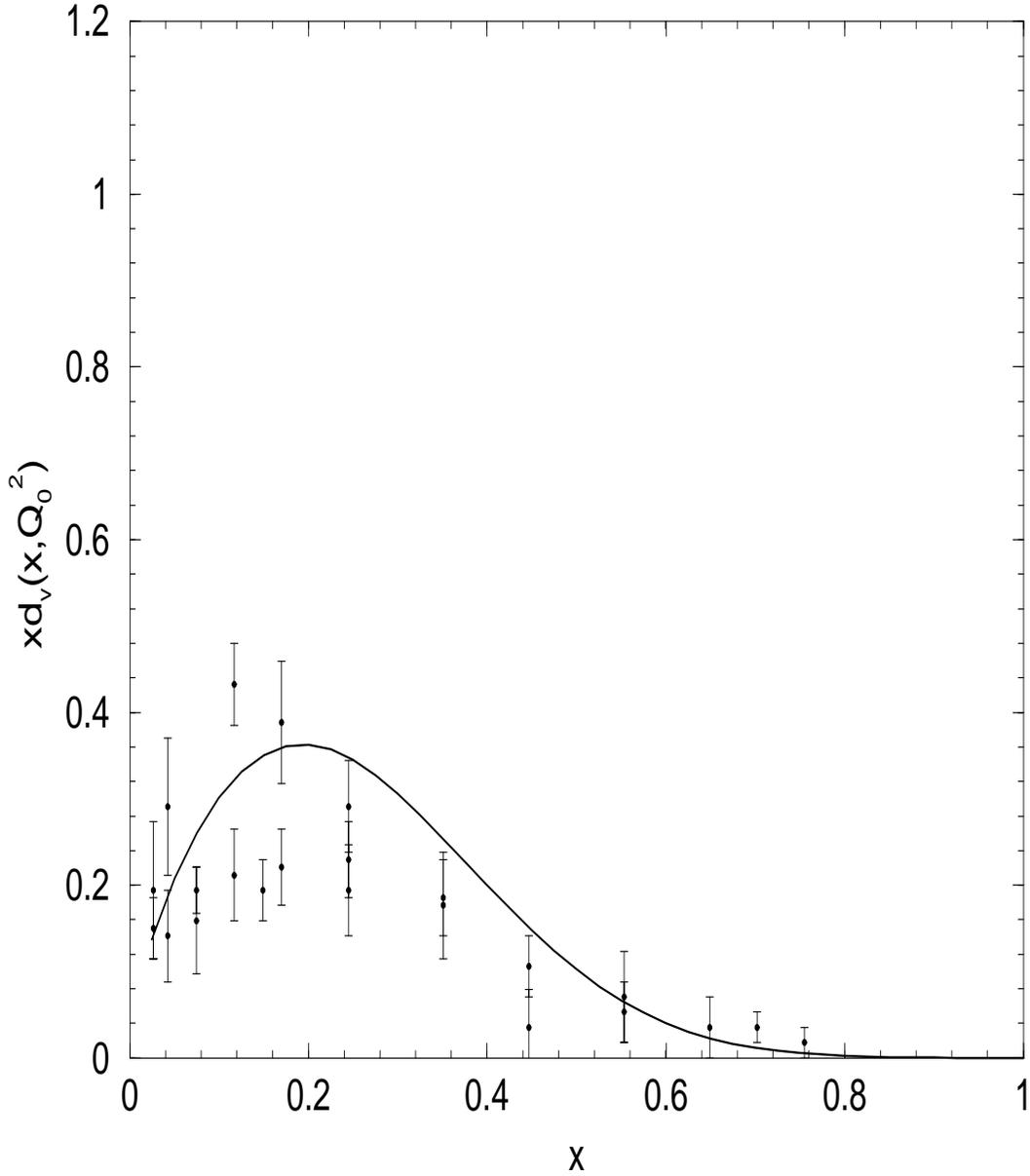}}

\caption{The QCD evolved result for $xd_v(x,Q^2)$ at
$Q_0^2=15GeV^2$ (solid line) is given in comparison with
the experimental data taken from T.Sloan {\it et~al} in Ref.[2] }

\end{figure}

\begin{figure}[t]

\protect\centerline{\epsfxsize=5.5in \epsfysize=6.5in \epsfbox{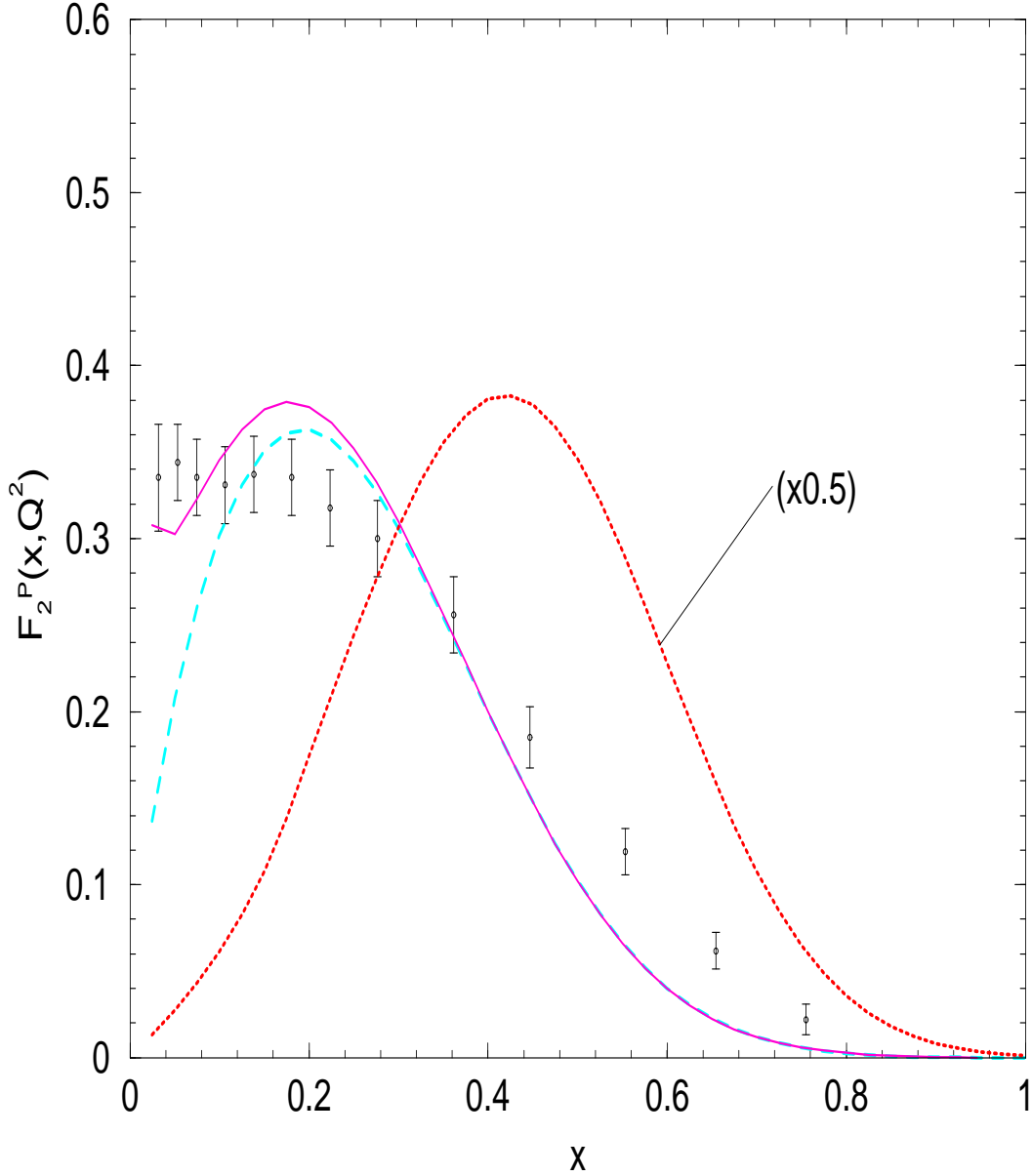}}

\caption{The calculated $[F_2^p(x,Q^2)]_{val}$ at $Q^2={\mu}^2=0.1GeV^2$
(dotted line) and its QCD evolved result at $Q_0^2=15GeV^2$ (dashed line).
$F_2^P(x,Q^2)$(valence+asymmetric sea: solid line) in comparison
with experimental data taken from R.G.Roberts.{\it et~al} in Ref.[1].}

\end{figure}

\begin{figure}[t]

\protect\centerline{\epsfxsize=5.5in \epsfysize=6.5in \epsfbox{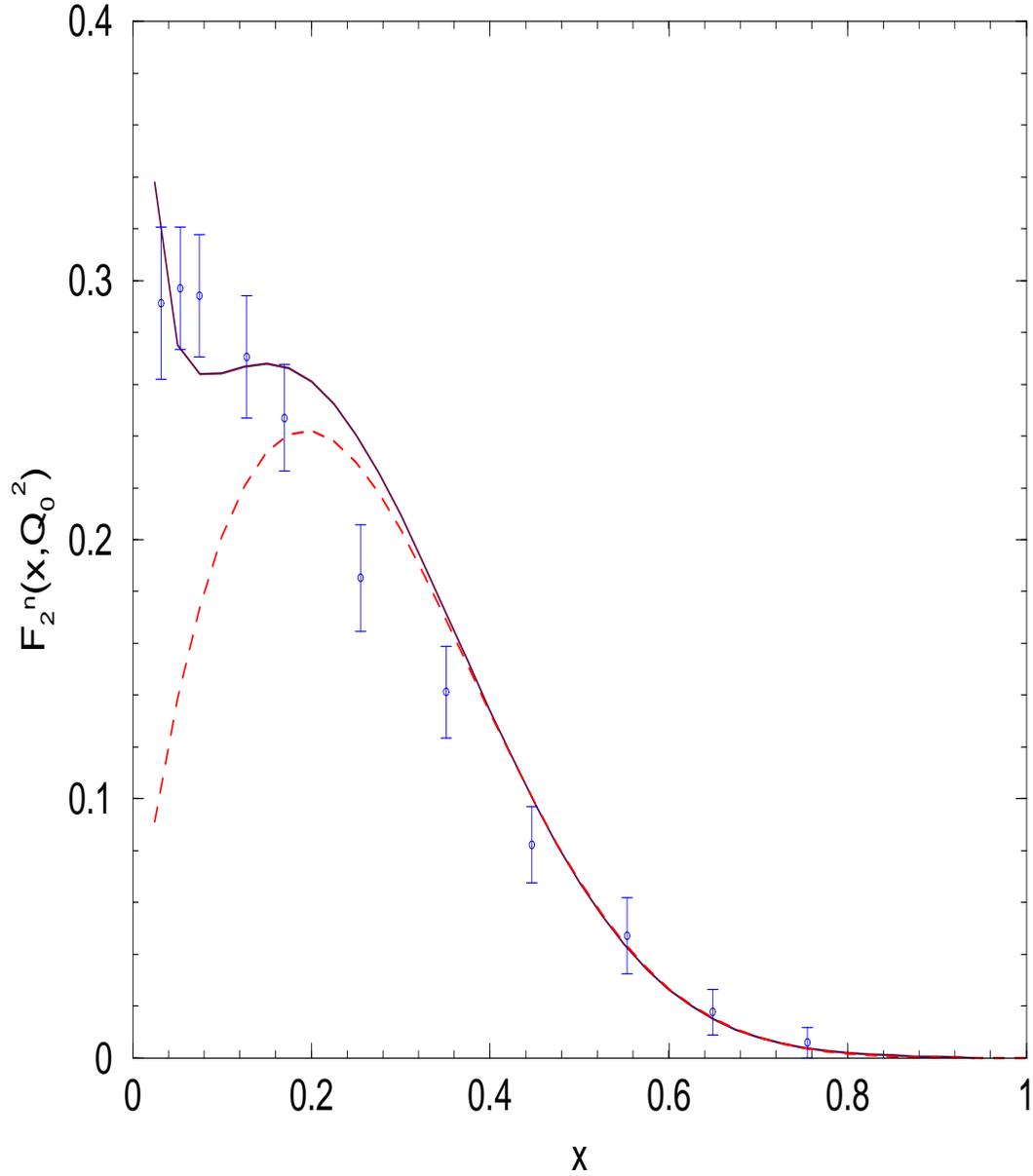}}

\caption{The QCD evolved result for $[F_2^n(x,Q^2)]_{val}$ at
$Q_0^2=15GeV^2$ (dashed line) and $F_2^n(x,Q^2)$
(valence+asymmetric sea: solid line) in comparison with data
from R.G.Roberts.{\it et~al} in Ref [1].}

\end{figure}

\begin{figure}[t]

\protect\centerline{\epsfxsize=5.5in \epsfysize=6.5in \epsfbox{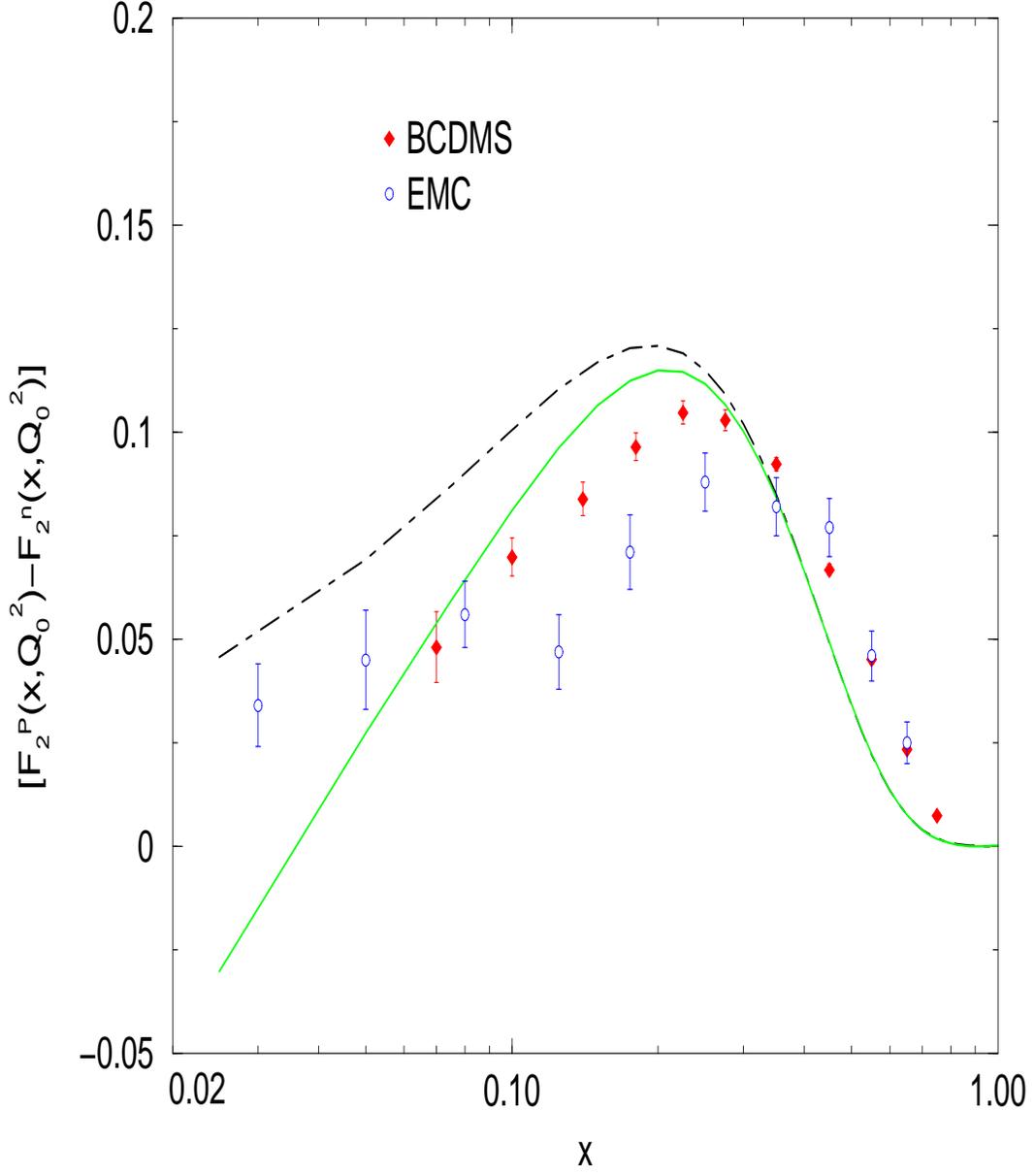}}

\caption{The QCD evolved result for $[F_2^p(x,Q_0^2)-F_2^n(x,Q_0^2)]$
(dot-dashedline-valence only; solid line- valence+asymmetric sea)
at $Q_0^2=15GeV^2$ compared with the data.(data is
over the $Q^2$-range of the experiments as per Ref.[2])}

\end{figure}

\begin{figure}[t]

\protect\centerline{\epsfxsize=5.5in \epsfysize=6.5in \epsfbox{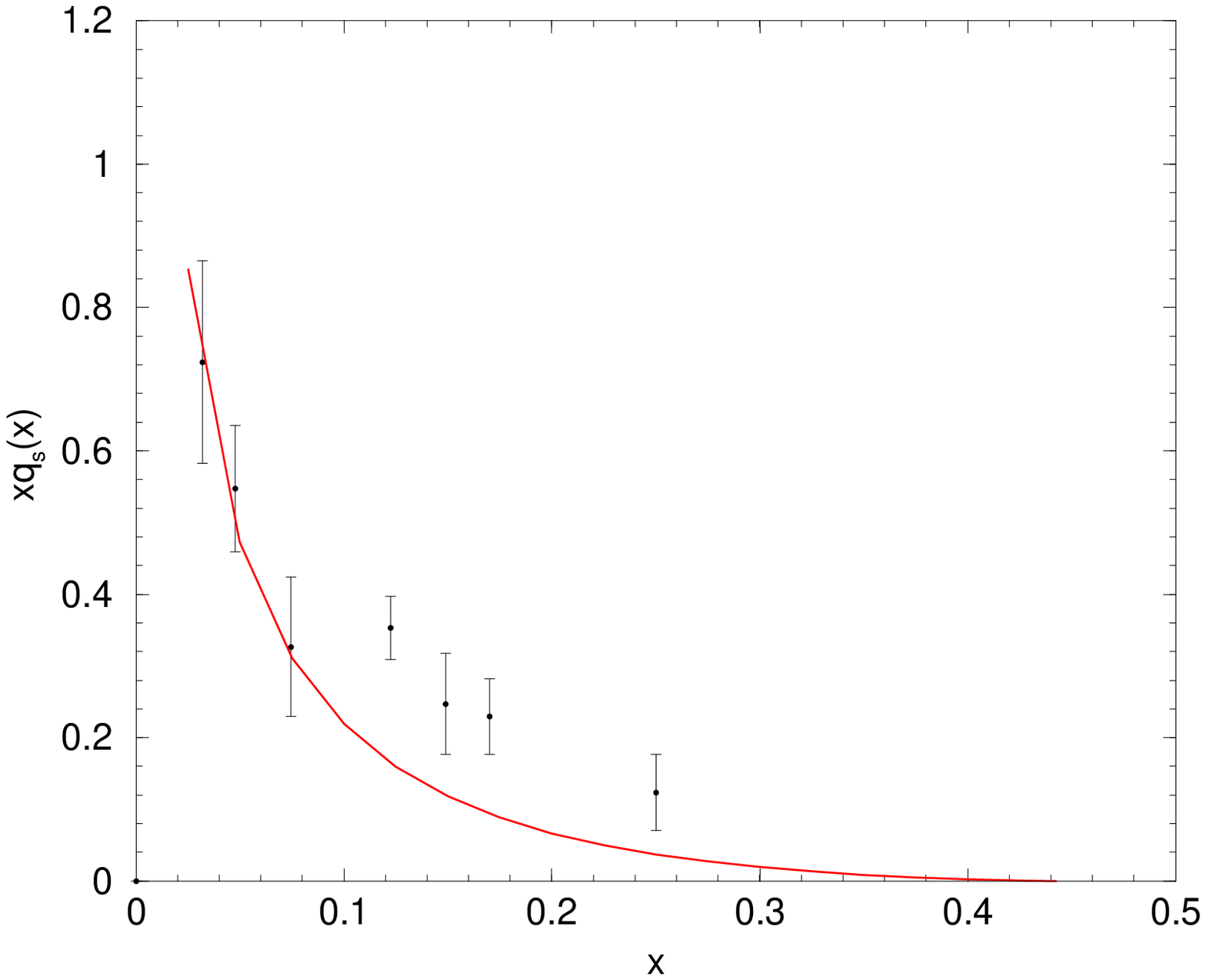}}

\caption{The dynamically generated $xq_s(x,Q^2)$(solid line) at $Q_0^2
=15GeV^2$, compared with the data from T.Sloan {\it et~al}
in Ref [2]. }

\end{figure}

\begin{figure}[t]

\protect\centerline{\epsfxsize=5.5in \epsfysize=6.5in \epsfbox{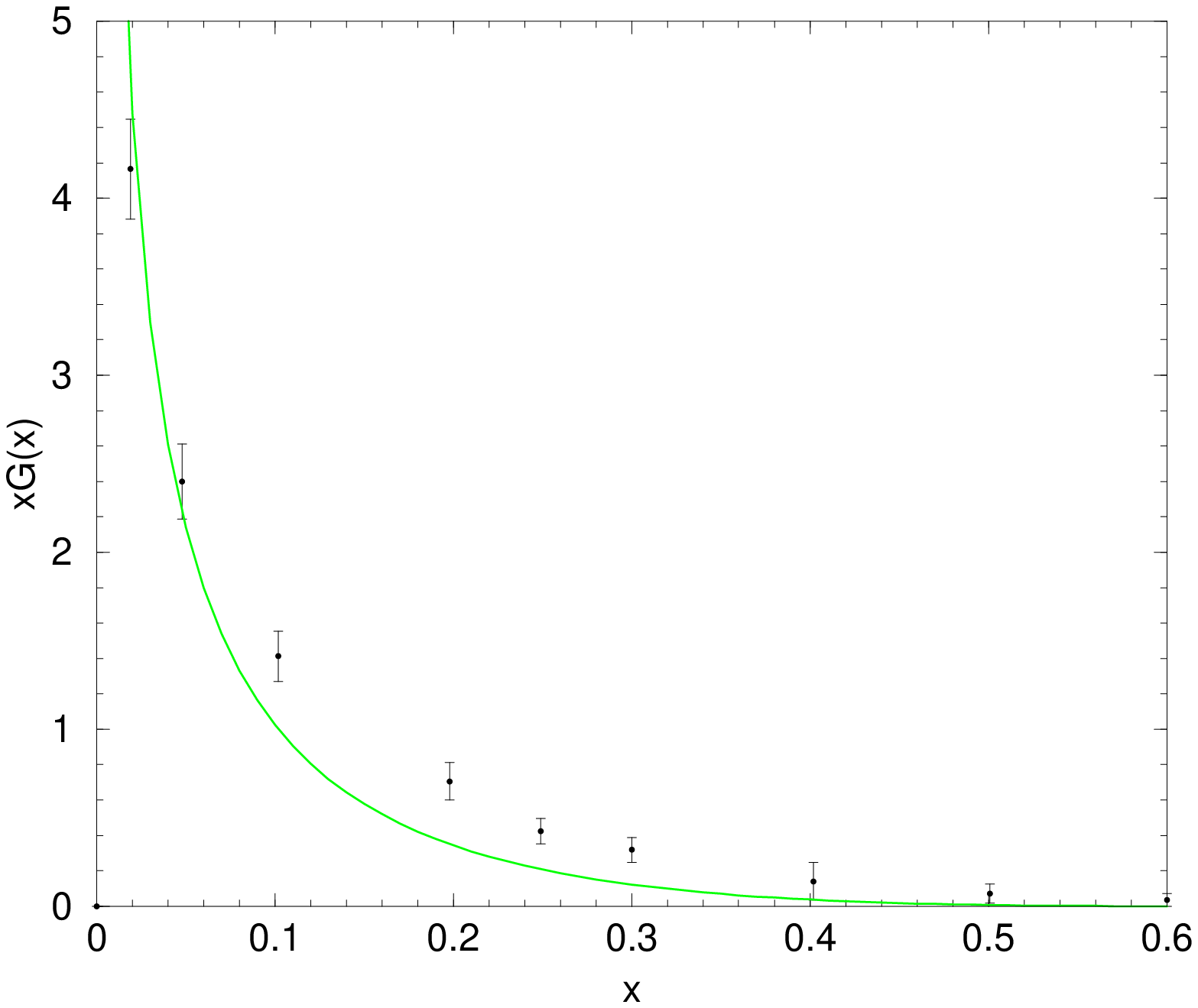}}

\caption{The dynamically generated $xG(x,Q^2)$(solid line) at $Q_0^2
=15GeV^2$, compared with the data from T.Sloan {\it et~al}
in Ref [2].}

\end{figure}


\begin{references}
\bibitem{callen} Callen Gross relation is satisfied in the model
here, since it can be shown that $W_{00}^{(S)}(x,Q^2)$ is
finite in the Bjorken limit as ;
$W_{00}^{S}={\frac{M}{4I_N}}[\int_{K_{+}}^{\infty}{\frac{dK}{K^2}}
{\rho}(K){\tilde {\rho}}(K)\{K+K_{+}(x)\}-(K_{+}{\longrightarrow}K_{-})]$
\end{references}
\end{document}